# Optimising Temporary Accommodation Placement Across London with AI-Powered SaaS in E-Governance Systems

Hankun He,[1,2] Jordan Richards,[1,3] Gopalakrishnan Netuveli,[1,2,] Kumar Aniket,[2] Ramya Pachatcharam,[1,2] Binta Ade-olusile,[1,2] Nathan Nagaiah,[1,4] and Matthew I Bellgard [1,3]

[1] UK Centre for AI in the Public Sector, London, United Kingdom

[2] Institute for Connected Communities, University of East London, London E15 4LZ, United Kingdom

[3] University of East London, London E15 4LZ, United Kingdom

[4] London Borough of Newham, London E16 2QU, United Kingdom

**Abstract**: Temporary accommodation has become a major fiscal and administrative pressure for English local authorities, particularly in London, where demand and costs have risen sharply. This paper documents the creation and use of DOMUS, a cloud-based, AI-enabled decision-support system built from scratch at the University of East London and customised for the needs of London Borough of Newham to support statutory Temporary accommodation placement. DOMUS integrates household case records, policy-constrained affordability and suitability rules, and live private-rental listings within a single governance-aligned workflow. The system combines transparent, rule-based filtering with large language model-assisted search to standardise the application of bedroom need, affordability thresholds, geographic preferences, and accessibility requirements, while preserving officer discretion and audibility. Household and property attributes are encoded into policy-consistent representations prior to AI-assisted ranking and explanation. A pilot deployment in Newham's secure environment evaluated operational performance relative to manual workflows. Results indicate substantial reductions in search time, improved adherence to key placement constraints, and high staff satisfaction, while maintaining statutory compliance and role-based accountability. Beyond TA, the paper frames DOMUS as replicable digital public infrastructure: a modular, cloud-native Software-as-a-Service architecture that can be deployed across other UK boroughs and adapted to other public administration tasks characterised by scarcity, rule-bound eligibility, and high stakes. The findings demonstrate the feasibility of scalable, ethically governed AI deployment in local government and contribute to debates on AI-enabled public value creation in e-governance.

## 1. Introduction

Temporary accommodation (TA) is a growing crisis in the UK, with councils spending over £1.6 billion annually on temporary accommodation alone [1]. In England, the number of households in temporary accommodation reached 131,140 on 31 March 2025, an increase of 11.8% from 31 March 2024 [2]. Households with children in temporary accommodation increased by 11.6% to 83,150, while single households increased by 12.0% to 47,990 over the same period [2]. These pressures are particularly acute in London, with 19.9 households in temporary accommodation per 1,000 households, compared with 2.8 per 1,000 in the rest of England [2].

This is especially evident in the London Borough of Newham (LBN), where as of October 2024 around 1 in 20 households were in temporary accommodation (6,662 households) [3]. Numbers in TA in Newham have grown by 145% since 2013, and costs rose from a mean nightly rate of £42.16 in April 2021 to a peak of £158.23 in November 2023 [3]. The council reported a £31 million overspend in 2024, underscoring the fiscal strain created by TA demand [3]. Behind every one of those cases is a decision — about where a family lives, what school a child can attend, and how long they remain in uncertainty. The council is legally required to find suitable accommodation for any household that is homeless, eligible, and in priority need. Under this level of demand and financial pressure, making those decisions quickly and consistently becomes increasingly difficult. This is the context in which DOMUS was developed and early-tested: to explore whether AI-assisted matching can help councils identify suitable placements more efficiently, while improving adherence to key placement requirements compared with manual workflows.

To address this challenge, the UK Centre for AI in the Public Sector, a partnership between the University of East London (UEL) and the LBN, developed DOMUS. DOMUS is a decision-support system to automate the matching of households in temporary accommodation to suitable private-rental properties. DOMUS reads registered household needs and safely scrapes live listings from external portals, then applies policy-aware checks, such as bedroom need, affordability, preferred areas, and accessibility, to identify closely matching properties. It enables faster, more equitable placements and reduces time in temporary accommodation. Piloted with Newham Council, the system demonstrates scalable, ethical AI deployment for housing allocation in the public sector. This paper presents the system architecture, pilot findings, and its potential to improve the speed, transparency, and fairness of housing services.

## 2. Literature Review

Research on AI-enabled decision support for temporary accommodation (TA) allocation remains limited. Most relevant evidence is therefore drawn from adjacent literatures on homelessness prediction, homelessness prevention, the allocation of scarce housing or homelessness services, and fairness in algorithmic public-sector decision making. Existing reviews suggest that this literature has expanded in recent years, but remains concentrated largely in North America, with relatively little peer-reviewed research directly addressing UK temporary accommodation workflows [4, 5].

The studies most relevant to TA allocation are those that examine AI-supported matching and prioritisation under multiple constraints. For example, [6] present a housing recommender system that combines listing ingestion, geospatial enrichment, and reinforcement-learning-based recommendation to identify budget-constrained housing options in Timis,oara. Although developed for student housing rather than homelessness or TA, the study demonstrates how AI systems can search across listings and rank options according to multiple constraints. Similarly, [7] examine the use of AI decision aids in homeless youth housing systems. Their study evaluates whether existing prioritisation rubrics predict later homelessness outcomes and argues for stronger human– AI collaboration in housing allocation decisions.

Related work also considers how scarce housing services can be allocated more efficiently. Using administrative homelessness-system data, [8] model counterfactual outcomes under different service assignments and formulate an optimisation problem for allocating scarce housing services in ways that reduce re-entry into homelessness. Their findings suggest that algorithmic allocation can improve system efficiency, but also highlight ethical trade-offs: some households may be assigned to interventions that increase their individual risk even where overall system-level outcomes improve. An important strand of the literature focuses on how such systems should be evaluated. [9] argue that domain-agnostic performance measures are insufficient for algorithmic homeless service allocation. Instead, they propose domain-specific evaluation metrics based on realistic "what-if" scenarios, emphasising that these systems should be assessed in terms of their practical consequences for service delivery and outcomes rather than predictive performance alone [9].

A further body of research highlights fairness, bias, and explainability as central concerns in high-stakes housing decisions. Reviews of homelessness-related algorithms show that only 15.7% of studies explicitly address fairness or bias, despite the sensitivity and potential consequences of these decisions [4]. This concern is reinforced by empirical work comparing different assessment tools. For instance, [10] compare the Vulnerability Index–Service Prioritization Decision Assistance Tool (VI-SPDAT), one of the most widely used assessment tools in the United States, with the Allegheny Housing Assessment (AHA), a newer administrative-data-based predictive tool. They find that the earlier tool exhibits racial bias and that, although the newer model aligned service decisions more closely with estimated risk, it did not automatically eliminate racial disparities in service rates. The persistent disparity is attributable to alternative assessment pathways and eligibility factors, including the use of Alt-AHA, a survey-based tool.

Beyond allocation itself, another strand of research extends these approaches to homelessness prevention. [11] show that machine-learning-informed prioritisation of households facing eviction, based on their risk of future homelessness, can improve the targeting of rental assistance. Their results suggest that such approaches can identify households who might be overlooked by more reactive systems while also meeting equity goals across race and gender. In a related study, [12] combine counterfactual machine learning with community input to develop more transparent prioritisation rules for subgroups such as families, youth, and people with comorbid health conditions. They argue that prioritisation frameworks should not only maximise the use of scarce resources but also explicitly address disparities associated with minoritisation and marginalisation.

Taken together, this literature suggests that AI has significant potential to support housing and homelessness systems by improving risk identification, strengthening prioritisation and matching under resource constraints, and increasing the consistency and transparency of decision support. At the same time, the strongest studies emphasise the importance of careful evaluation, fairness-aware design, explainability, and continued human oversight. There remains a clear gap in the literature on caseworker-facing, AI-enabled systems designed specifically for UK temporary accommodation workflows, particularly tools that integrate affordability and eligibility constraints, property identification and ranking, and explainable matching within operational allocation processes.

## 3. Methodology

### 3.1. Question Identification

The above considerations motivate a set of research questions of broader relevance to AI deployment in the public sector, with a focus on supporting TA services in high-cost settings such as London, where vulnerable households face acute affordability constraints. The study is guided by the following questions:

1. To what extent does DOMUS reduce the time required to identify suitable property matches compared with the manual workflow?
2. Does AI-assisted matching improve adherence to key placement constraints (e.g., affordability, bedroom size, zone preference, and accessibility requirements) relative to current practice?
3. How do frontline officers perceive the usability, transparency, and trustworthiness of DOMUS, including the usefulness of explanations and audit logs?
4. What operational and governance considerations affect the feasibility of deploying DOMUS ethically and at scale within local authority workflows?
5. Does DOMUS reduce variability in matching decisions across officers by standardising the application of placement rules?

These questions are addressed in the remainder of the paper.

### 3.2. Data Preprocessing

DOMUS transforms administrative household records and private-market listings into a policy-aligned representation for matching. Household needs and property attributes are encoded as compact feature codes: short alphanumeric strings that capture non-negotiable constraints such as bedroom requirement, affordability, preferred zone, and accessibility needs. This representation supports consistent, auditable rule-based filtering before any AI-assisted retrieval is applied. Anonymised household records were provided by the London Borough of Newham and included structured fields covering bedroom need, affordability limits, zone preference, and accessibility flags. Live property listings are collected from third-party platforms, capturing rent, bedroom count, and postcode-level location. External lookup tables were used to attach Local Housing Allowance (LHA) rates and postcode-to-zone mappings to enable policy-consistent affordability checks and geographic standardisation. Service-access considerations were incorporated to reflect continuity of essential services, including schools, GP practices, hospitals, and support networks. Where applicable, these requirements were operationalised as measurable indicators (e.g., distance or travel-time constraints) and used alongside housing constraints to support filtering and prioritisation. Each record was encoded into a feature code ID string (e.g., N200358A10), enabling rule-based filtering on affordability, bedroom size, zone, and accessibility [4]. For AI-assisted search, the same codes were rendered into natural-language prompts (e.g., "Find a 2-bedroom property in zone A1 under £358"), supporting retrieval and explanation while preserving the underlying rule-based policy logic.

### 3.3. Nature of the Algorithm

The design goal is to reduce manual effort in searching, applying affordability constraints, and following up on shortlisted properties, while keeping key actions and outcomes visible to caseworkers throughout the allocation cycle. At a logical level, DOMUS comprises a browser-based frontend delivered through an Edge/CDN layer, an identity and access layer for secure sign-in, an API gateway providing controlled access to backend functions, an application services layer for household, search, matching, and administrative operations, and a persistence layer supporting operational data, cache, and audit/telemetry records. The AI-enabled search component combines deterministic filtering with model-assisted interpretation of officer inputs and case information through a Claude LLM service. It reviews and analyses household information and user inputs provided through a chat-based interface before returning filtered or sanitised results to the client. The engine provides clear reasons for each recommendation and answers follow-up questions from officers. The result is a simple, secure, and scalable AI tool for temporary accommodation teams.

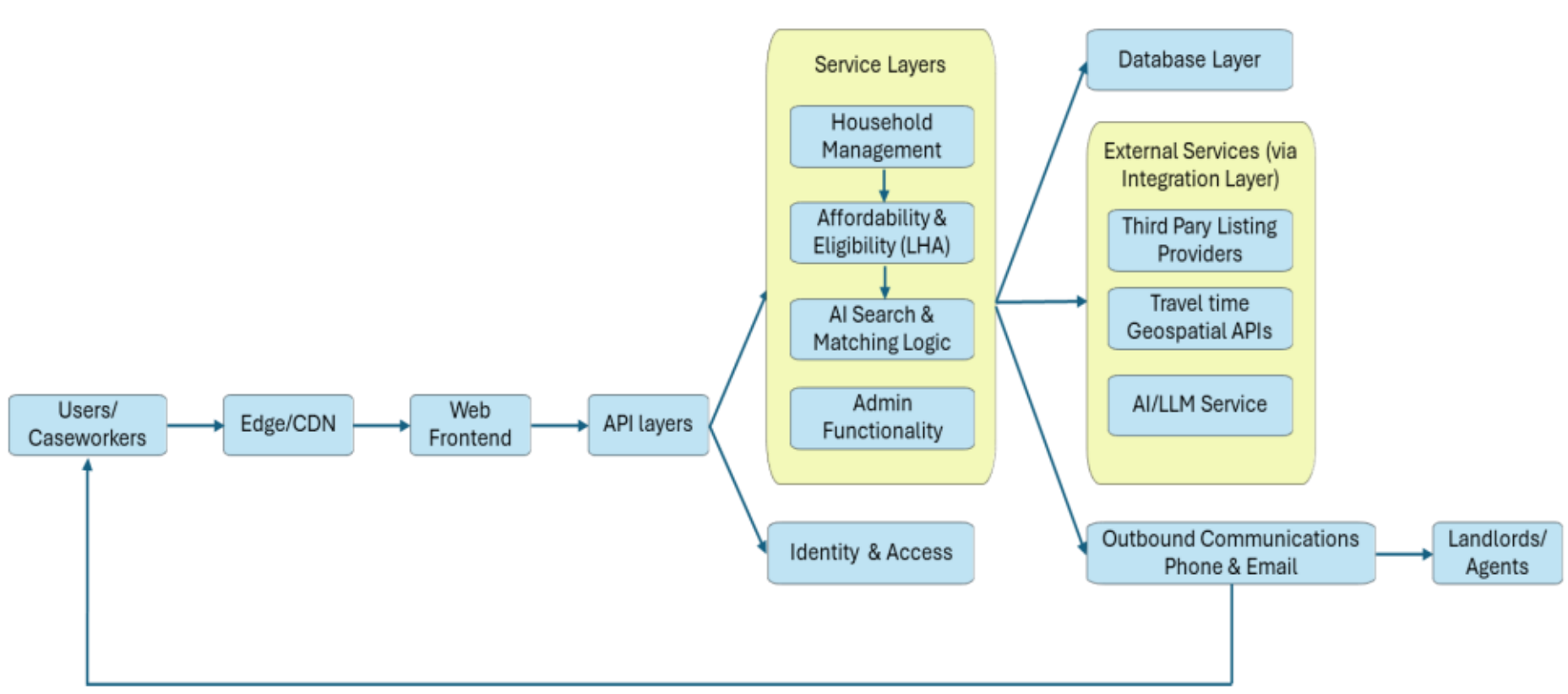

**Fig. 1.** Logical architecture of DOMUS. Users/Caseworkers access a static web frontend via an Edge/CDN and invoke backend capabilities through API layers secured by Identity & Access. Core service layers (Household Management, AI Search & Matching Logic, and Admin Functionality) persist operational state in a database layer, integrate with external services via an integration layer (listing providers, travel-time/geospatial APIs, and optional model-based services), and support outbound communications (phone/email) with Landlords/Agents, with notifications looping back to Users/Caseworkers.

Figure 1 illustrates the high-level architecture of DOMUS, which supports TA allocation workflows for Users/Caseworkers. End users access the system through a Web Frontend delivered via an Edge/CDN layer, which reduces latency and improves responsiveness for commonly requested static assets. The frontend communicates with backend functionality through API layers, providing a uniform entry point for client requests and enabling consistent application of security and governance controls.

Backend functionality is structured as a set of service layers comprising: (i) Household Management for maintaining case and household records; (ii) an Affordability & Eligibility (LHA) component that derives affordability constraints and applies eligibility filtering based on household attributes and preferred areas; (iii) AI Search & Matching Logic for identifying suitable properties that satisfy household needs and LHA-derived affordability constraints; and (iv) Admin Functionality for administrative and operational management tasks. The service layer persists and retrieves operational state through a database layer, allowing the system to support repeatable allocation cycles, case continuity, and data consistency across user sessions. Where externally sourced listings are used to support search and matching, these are incorporated through an integration layer that supports compliance-aware ingestion, data normalisation, and governance controls before they enter allocation workflows.

Authentication and authorisation concerns are separated from the core logic through an Identity & Access component integrated with the API tier. This separation supports a consistent access-control model across services and reduces coupling between security mechanisms and domain functionality. The resulting interaction pattern follows a standard web SaaS flow: authenticated users invoke API endpoints that route requests to the relevant service components, which in turn read and write to the database layer as needed.

DOMUS interfaces with external dependencies through an External Services (Via Integration Layer) block. This includes Third Party Listing Providers, Travel time Geospatial APIs, and an AI/LLM Service. Abstracting external dependencies behind an integration layer provides a stable boundary for handling variability in third-party systems, supports policy enforcement (e.g., request mediation and governance), and limits direct coupling between internal service logic and external interfaces.

Operationally, the architecture includes an Outbound Communications (Phone & Email) component that enables the platform to initiate contact with Landlords/Agents as part of the accommodation engagement process. Notifications and updates can be routed back to Users/Caseworkers, supporting a closed-loop workflow in which shortlists, contact attempts, and subsequent updates remain coordinated within the platform rather than being managed entirely outside the system.

### 3.4. Functionality

DOMUS provides end-to-end support for temporary accommodation (TA) allocation workflows, combining case management, cross-platform property retrieval, AI-assisted matching, and operational coordination with landlords/agents. Its functionality can be summarised across the following capability areas:

a) Household and case management. DOMUS maintains structured household profiles containing attributes used in allocation decisions (e.g., household composition, bedroom requirements, and preferred areas). These profiles parameterise searches and support repeatable allocation cycles by ensuring consistent criteria are applied across sessions.
b) Live Cross-Platform Data. DOMUS aggregates live property listings from Zoopla, Rightmove, and OnTheMarket using a secure and policy-compliant web data acquisition pipeline, providing access to thousands of current listings across the M25.
c) Nearby Amenities Candidate properties can be enriched with contextual information, including travel-time/geospatial evidence and proximity to relevant amenities (e.g., schools, primary care services, and hospitals) with links to authoritative sources.
d) AI-assisted search and matching. DOMUS supports shortlisting through a matching component that filters and ranks properties against household needs and user-defined constraints (e.g., region, bedroom count, and source platform). An embedded assistant supports interactive refinement by answering follow-up questions and providing explanations of recommendations to improve transparency and user oversight.
e) Affordability checks. To support affordability screening during shortlisting, DOMUS retrieves and applies LHA rates based on household details and preferred locations.

f) Real-Time flagging System. Caseworkers can track and update property engagement states (e.g., available, unavailable, not suitable, reserved), enabling a consistent operational view during follow-up.
g) Operational communications. DOMUS supports outbound communications, allowing caseworkers to send selected property lists by email and to initiate phone/email contact with landlords/agents from within the platform. Updates and notifications can be routed back to caseworkers to support closed-loop coordination.
h) Feedback & Issue Loop. A lightweight feedback mechanism enables users to report issues and submit feature requests to support iterative improvement.
i) Administration. Administrative functionality supports operational management tasks (e.g., overseeing records and configurations). The interface provides an inline property summary view within the main workflow, enabling rapid inspection of key attributes without leaving the current screen.

## 3. Ethical & Legal Compliance

DOMUS operates under clear statutory authority and incorporates ethical safeguards, strengthening its position against potential legal challenge. This section outlines the statutory basis for deployment and the key ethical measures implemented.

### 3.1. Statutory Basis

**Table 1.** DOMUS operates under clear statutory authority derived from the Housing Act 1996.

| Section | Duty | DOMUS Application |
|---|---|---|
| 189B Relief Duty | *"Housing authorities have a duty to take reasonable steps to help any eligible applicant who is homeless secure accommodation"* | Searching live rental listings across multiple portals is a "reasonable step" to identify available properties. |
| 193(2) Main Housing Duty | *"The housing authority has a duty under section 193(2) to ensure that accommodation is available for their occupation"* | Discharging this duty requires access to current market availability - properties let quickly and data must be live. |
| 206 Suitability of Accommodation | *"Where a housing authority discharges its functions to secure that accommodation is available for an applicant, the accommodation must be suitable"* | Matching households to suitable properties (bedrooms, accessibility, location, affordability) requires comprehensive, up-to-date listing |

DOMUS supports the LBN in discharging its legal duties to assist residents in housing need by enabling efficient searches for suitable accommodation in the private rental market. The public interest in supporting vulnerable residents, combined with the system's non-commercial purpose and minimal impact on data sources, provides a strong justification for its operation.

### 3.2. Ethical Safeguards

**Table 2.** Technical measures ensure minimal burden on data sources from web scraping.

| Practice | Implementation | Purpose |
|---|---|---|
| Rate Limiting | 1-2 second delays between requests | Prevent server overload |
| Public Data Only | No login/authentication | Only access publicly visible listings |
| Non-Commercial | Council use only | No data resale or monetization |
| Data Freshness | 90-day automatic data deletion | No indefinite data hoarding |
| Minimal Collection | Essential fields only; no bulk downloading or data hoarding | Address, price, bedrooms, images |
| No Redistribution | Internal use only | Data stays within Council systems |

Overall, DOMUS is underpinned by statutory authority, demonstrable public benefit, ethical implementation, and limited commercial impact.

## 4. Performance Evaluation

DOMUS was piloted with the LBN in early 2025 to assess whether AI-assisted property matching could improve placement speed and match quality relative to the existing manual workflow. Housing officers took part in a live simulation using anonymised household records and real property listings. Following a short onboarding session, participants used DOMUS to retrieve cases via household reference IDs and to run searches using natural-language prompts. User experience and perceived utility were captured through a brief survey and informal interviews.

DOMUS was deployed entirely within the council's secure computing environment. Officers accessed the tool through a browser-based interface, reviewed ranked recommendations accompanied by LLM explanations, and shared shortlisted matches via email. The pilot was designed to fit within existing operational practices and did not require major system changes.

Overall, the pilot demonstrated clear efficiency gains, improved match accuracy, and high staff satisfaction. Table 3 summarises the key performance metrics, integrating quantitative results and qualitative feedback from frontline officers, and compares manual practice with the AI-assisted approach. The average time to identify suitable matches fell from approximately 2–3 hours to under one minute. Compliance with key constraints, including affordability, bedroom size, and zone preference, increased from around 65% under the manual process to approximately 85–90% with DOMUS. In addition, match results could be communicated immediately via email, rather than with a 24-hour turnaround. Staff satisfaction averaged 4.5 out of 5, reflecting strong approval across the housing team.

Frontline caseworkers highlighted three recurring strengths. First, speed was transformative: substantially faster searches enabled officers to handle more cases each day and devote more time to complex cases. Second, transparency improved through LLM-generated explanations, which enhanced officers' confidence in the recommendations. Third, usability was a major advantage, with an intuitive interface requiring minimal training.

**Table 3.** Summary of pilot results comparing manual and AI-assisted matching.

| Performance Metrics | Manual Process | DOMUS |
|---|---|---|
| Avg. time to match | 2–3 hours | < 1 minute |
| Constraint compliance | ~65% | 85–90% |
| Email delivery time | ~24 hours | Instant (1-click) |
| Officer satisfaction score | N/A | 4.5/5 |

These findings draw on the development of DOMUS, early trial use, and demonstration in council settings. They show the tool's practical potential to support more transparent, consistent, and data-informed temporary accommodation decisions. The next-phase Newham beta pilot will extend this work and strengthen the evidence base further. The findings can be summarised in the following key points:

1. Integrates key stages of TA allocation within a single platform — bringing together case management, affordability and eligibility checks (e.g. LHA rates), cross-platform property listing retrieval, and decision support in one cloud-based, AI-enabled system.
2. Delivered measurable operational and governance benefits in the Newham pilot, with wider relevance for public sector housing services. Matching times fell from several hours to under a minute, increasing throughput and allowing staff to focus on more complex cases.
3. Improved compliance with key placement rules, rising from around 65% to approximately 90%, and demonstrating the practical feasibility of scalable and ethical AI deployment in public sector housing allocation.
4. Strengthened fairness, transparency, and auditability through rule-based filtering, explainable recommendations, and action logging, helping make allocation decisions more consistent and accountable.

## 5. Discussion

The DOMUS pilot delivered measurable operational and governance benefits for Newham Council, with wider relevance to public sector housing services. Matching times fell from several hours to under a minute, supporting higher throughput and enabling staff to focus on complex cases. Rule-based filtering strengthened fairness and consistency, while explainable recommendations and action logging improved transparency and auditability. The approach may also reduce costs by shortening stays in temporary accommodation. These findings imply:

1. Transforms statutory “suitability” criteria into measurable indicators (distance/travel-time to school, GP, support networks), making decisions more consistent across officers and boroughs.
2. Quantifies trade-offs under scarcity, helping leaders compare scenarios (e.g., cost vs. disruption) and identify thresholds where out-of-area placement becomes disproportionate.
3. Improves monitoring of out-of-area practice, enabling boroughs and London system partners to track where and why placements are being made outside areas and to target reforms accordingly.
4. Supports better value for money in a high-cost environment, where TA is a major and rising budget pressure and councils face increasing strain from homelessness duties.
5. Helps shift from crisis management to prevention-aligned operations, by enabling earlier identification of households at risk of severe disruption and prioritising placements that preserve stability (schooling, health care, caring).

AI will not solve the housing shortage, but it can fundamentally improve how we manage it — making decisions more consistent, transparent, and aligned with human need.

## 6. Conclusions and Future Outlook

DOMUS demonstrates how a cloud-based, AI-enabled SaaS platform can support temporary accommodation (TA) allocation by unifying case management, affordability/eligibility constraints (including LHA-derived thresholds), cross-platform listing retrieval, and decision support within a single workflow. Household records feed an explicit affordability component, which informs AI-assisted search and matching alongside other requirements and contextual signals (e.g., location and travel-time evidence). A secured service-oriented backend with managed persistence and identity and access control provides consistent governance, while edge delivery maintains responsive user access. Beyond shortlisting, DOMUS operationalises engagement through real-time status flagging and integrated outbound communications (email and phone) to landlords/agents, with updates routed back to caseworkers to close the loop. A built-in feedback and issue mechanism supports continuous improvement. Future work will strengthen the AI search logic, improve explanation robustness, enhance data-quality controls for live listings, and extend bias mitigation and deployment evaluation through pilot trials in high-stakes public-sector settings.

## Data Availability Statement

The implementation of the method used in the paper is available from the authors upon reasonable request, subject to approval by the University of East London and the London Borough of Newham.

## Acknowledgments

We thank Shumaila Ahmed, and Daniel Okoroma for helpful discussions related to the machine learning and AI aspects. We also thank Joanna Hansfield and Hannah Chang for valuable insights that helped the development of DOMUS.

## References

1. Davies, G. The effectiveness of government in tackling homelessness. National Audit Office (2024). last accessed 2026/02/24.
2. Ministry of Housing, Communities and Local Government. Statutory homelessness in England: January to march 2025 (2025). last accessed 2026/02/24.
3. London Borough of Newham. Mayor speech to introduce debate - full council 21 oct 2024 (2024). last accessed 2026/02/24.
4. Moon, E. S.-Y. & Guha, S. A human-centered review of algorithms in homelessness research. In Proceedings of the 2024 CHI conference on human factors in computing systems, 1–15 (2024).
5. Stängl, L. A. et al. Homelessness prediction models in high-income countries: a scoping review. BMC Public Health 25, 3964 (2025).
6. Nicula, A.-S., Ternauciuc, A. & Vasiu, R.-A. A smart housing recommender for students in Timișoara: Reinforcement learning and geospatial analytics in a modern application. Applied Sciences 15, 7869 (2025).
7. Chan, H., Rice, E., Vayanos, P., Tambe, M. & Morton, M. Evidence from the past: AI decision aids to improve housing systems for homeless youth. In AAAI Fall Symposia, 149–157 (2017).
8. Kube, A. R., Das, S. & Fowler, P. J. Fair and efficient allocation of scarce resources based on predicted outcomes: Implications for homeless service delivery. Journal of Artificial Intelligence Research 76, 1–44 (2023).
9. Qi, W. & Chelmis, C. Evaluating algorithmic homeless service allocation. Journal of Computational Social Science 6, 59–89 (2023).
10. Cheng, L., Drayton, C., Chouldechova, A. & Vaithianathan, R. Algorithm-assisted decision making and racial disparities in housing: A study of the Allegheny Housing Assessment Tool. In Proceedings of the AAAI/ACM Conference on AI, Ethics, and Society (AIES 2024) (2024).
11. Vajiac, C. et al. Preventing eviction-caused homelessness through ML-informed distribution of rental assistance. Proceedings of the AAAI Conference on Artificial Intelligence 38, 22393–22400 (2024). URL: https://ojs.aaai.org/index.php/AAAI/article/view/30246

12. Kube, A. R., Das, S. & Fowler, P. J. Community- and data-driven homelessness prevention and service delivery: Optimizing for equity. Journal of the American Medical Informatics Association 30, 1032–1041 (2023).
13. Xu, L., Skoularidou, M., Cuesta-Infante, A. & Veeramachaneni, K. Modeling tabular data using conditional GAN. Advances in neural information processing systems 32 (2019).

## Author Contributions

Conceptualization - M.B., G.N., and N.N.; Methodology - J.R., H.H., G.N., R.P., and B.A.; Visualisation - H.H. and J.R.; Writing original draft - H.H., J.R., G.N., K.A, and B.A.; All authors revised the manuscript and approved the final submitted draft.

## Competing Interests

The authors declare no competing interests.